\documentclass[american,aps,pra,reprint,tightenlines,superscriptaddress,twocolumn,twoside]{revtex4-1}
\pdfoutput=1

\usepackage[unicode=true,pdfusetitle, bookmarks=true,bookmarksnumbered=false,bookmarksopen=false, breaklinks=false,pdfborder={0 0 0},backref=false,colorlinks=false] {hyperref}
\hypersetup{ colorlinks,linkcolor=myurlcolor,citecolor=myurlcolor,urlcolor=myurlcolor}
\usepackage{braket,colortbl,amsthm,amsmath,amssymb,dsfont}
\definecolor{myurlcolor}{rgb}{0,0,0.7}
\usepackage{graphics,graphicx,relsize}
\usepackage{color}
 \usepackage{XCharter}
 \usepackage[T1]{fontenc}
\usepackage{mathptmx}
\usepackage[left=1.65cm,right=1.65cm,top=2.45cm,bottom=2.45cm]{geometry}
\usepackage[caption=false]{subfig}
 
\theoremstyle{plain}

\def\bea{\begin{eqnarray}}
\def\eea{\end{eqnarray}}
\def\ba{\begin{array}}
\def\ea{\end{array}}

\def\beq{\begin{equation}}
\def\eeq{\end{equation}}

\DeclareMathOperator{\Tr}{Tr}

\usepackage[normalem]{ulem}
 
\begin{document}
\title{Detecting genuine multipartite entanglement in three-qubit systems
with eternal non-Markovianity}

\author{Ankit Vaishy}
\affiliation{Center for Computational Natural Sciences and Bioinformatics,} 

\author{Subhadip Mitra}
\email{subhadip.mitra@iiit.ac.in}
\affiliation{Center for Computational Natural Sciences and Bioinformatics,} 
\affiliation{Center for Quantum Science and Technology,} 

\author{Samyadeb Bhattacharya} 
\email{samyadeb.b@iiit.ac.in}
\affiliation{Center for Security Theory and Algorithmic Research,
International Institute of Information Technology, Gachibowli, Hyderabad 500 032, India}
\affiliation{Center for Quantum Science and Technology,} 

\begin{abstract}
\noindent 
We devise a novel protocol to detect genuinely multipartite entangled states by harnessing quantum non-Markovian operations. We utilize a particular type of non-Markovianity known as the eternal non-Markovianity to construct a non-complete positive map to filter out the bi-separable states and detect genuine multipartite entanglement. We further propose a witness operator to detect genuinely multipartite entangled states experimentally based on this theory. Our study sheds light on a 
hitherto unexplored connection between entanglement theory and quantum non-Markovianity.
\end{abstract}
\maketitle

\section{Introduction}
\noindent
Entanglement is a fascinating feature, typical and ubiquitous to the science of quantum information and computation. Over the last few decades, we have seen entanglement, or more precisely bipartite entanglement, playing an unparalleled role in various quantum information-theoretic tasks~\citep{ekert91,bennett92,bennett93}. Multipartite entanglement comes with an even richer structure and a whole new set of features providing information processing advantages in quantum communication, simulation, or metrology \citep{multi1,multi2,multi3,multi5,multi6,multi7,multi8}. It is therefore essential to construct methods to identify genuinely multipartite entangled  states~\citep{wit6,guhne09,brukner04}. However, detecting entanglement of an arbitrary quantum state is a problem of significant complexity~\citep{gurvits,sevag}. Thus, witnessing entanglement by using versatile experimentally-feasible tools~\citep{wit1,wit2,wit3,wit4,wit5,wit6,guhne09,wit8,wit9,wit10} is a challenging task. In the case of multipartite entanglement, the challenge enhances significantly and demands greater interest from the scientific community.

The most direct way to detect a multipartite entangled state is to perform a complete state tomography, a challenging procedure for a few reasons. First, state tomography needs knowledge of the state parameters, which can be highly costly for a many-dimensional quantum state. Second, it is computationally hard to determine the separability classes from these complex data sets. Therefore, it is desirable to develop detection procedures that do not require complete knowledge of the quantum states. One such process, of course, is the construction of entanglement witnesses. However, since multipartite entanglement is still a complex and less-understood phenomenon without any canonical theory, it is challenging to construct witness operators for them. Thus, it is imperative to study and build different protocols of multipartite entanglement detection. In this paper, we propose one such protocol to detect multipartite entanglement by utilizing the well-known phenomenon of non-Markovianity in open quantum systems~\citep{rhp1,blp1} and the theory of positive maps.


The positive maps that are not completely positive can detect bipartite entanglement quite effectively, Partial Transposition~\citep{wit1} being a well-known example. Unfortunately, this technique cannot be directly extended to the case of multipartite entanglement, where diverse separability structures complicate the detection problem. Although semi-definite programming can be used to construct witnesses to certify genuine multipartite entanglement (GME)~\citep{guhne11,guhne15}, unlike the bipartite case, those witnesses have no correspondence with positive maps that can, in turn,  detect GME. Nonetheless, it has been shown that upon careful construction, positive maps can be useful for devising a method to certify GME \citep{sengupta14,huber17}. Harnessing this procedure, we construct a novel GME detection criteria based on the positive maps arising from a specific type of non-Markovian (NM) quantum operations.

The theory of open quantum systems provides us with a powerful tool to study system-environment interactions 
in various irreversible phenomena~\citep{alicki,breuer}.
Through the last decade or so, much effort has been made to characterize the quantum analog of NM evolution~\citep{rivas1,breuerN,alonso,blp1,rhp1,bellomo,arend,samya1,samya2,samya3,wolf1,nm1,nm2,nm3,nm4,nm5,nm6,samya4,samya5,samya6,samya7}. 
A physically well-understood (though it may not be exhaustive) way to do that is to consider all quantum dynamics that are not divisible in the set of NM evolution. Although it is still under scrutiny whether divisible NM operations exist, it is firmly established that indivisible operations are indeed NM. A divisible operation $\Phi(t_2,t_1)$ can be stated as 
\beq\label{1}
\Phi(t_2,t_1)=\Phi(t_2,t^\prime)\circ\Phi(t^\prime,t_1)
\eeq 
for all $t^\prime$, $t_1$, and $t_2$ such that  $t_1\leq t^\prime\leq t_2$. When an operation $\Phi_{\mathcal{N}}(t_2,t_1)$ is not divisible, then 
for at least some intermediate time ($t^\prime$), the map $\Phi_{\mathcal{N}}(t^\prime+\Delta, t^\prime)$ would not be completely positive. Interestingly, there exists a special type of NM operations called \emph{eternal} NM operations~\citep{hall14} that can produce positive but not completely positive maps in some intermediate timespan. If carefully executed, the eternal NM operations can, in turn, legitimately detect entangled quantum states. In this paper, we use such a quantum operation to construct a positive map to establish a protocol for detecting GME states. We also propose a GME identifier, which can be useful in detecting such entangled states experimentally.

The plan of the paper is as follows. In the next section, we elaborate on the concept of eternal NM evolution and how it can lead to positive maps in intermediate timesteps for detecting entangled states. We then establish the protocol of multipartite entanglement detection by manipulating such a particular type of NM operation in Section~\ref{sec:detection}. We further propose the witness to detect GME. Finally, we conclude with some possible implications.

\section{Eternal non-Markovianity and bipartite entanglement detection}
\noindent
To illustrate the concept of indivisible NM quantum operations, let us first restrict ourselves to the set of operations having Lindblad type generators~\citep{lindblad,gorini}. These have the form, 
\begin{align}
\hspace{-0.1cm}\dot{\rho}(t)=&\mathcal{L}_t(\rho(t))\nonumber\\
=&\mathlarger{\mathlarger{\sum}}_{i=1}^{n}
\Gamma_{i}(t)\left[L_{i}\rho(t)L_{i}^{\dagger}-\frac{1}{2}L_{i}^{\dagger}L_{i}\rho(t)-\frac{1}{2}\rho(t)L_{i}^{\dagger}L_{i}\right]\label{1a}
\end{align}
where $\Gamma_{i}(t)$ and $L_{i}$ are the Lindblad coefficients and operators respectively.
If the operation is divisible, then  $\Gamma_{i}(t)\geq 0,~ \forall i, t$ \citep{rhp1}. For indivisible NM operations, $\Gamma_{i}(t) < 0$ for some $i$ at some instant of time $t$. It was initially perceived~\citep{rhp1} that the condition $\int_0^T\Gamma_{i}(t)dt\geq 0~ (\forall i, T)$ must always hold to ensure complete positivity of the total dynamics. However, it was later established that for some legitimate quantum operations, one or more  Lindblad coefficients could be negative~\citep{hall14}, giving rise to the phenomenon of eternal NM. Here, we consider the simplest of such an operation which is fundamentally a qubit depolarizing operation.

Let us consider the following operation on a qubit expressed by the density matrix 
\[
\rho(t)=\left(\begin{matrix}
\rho_{11}(t)&\rho_{12}(t)\\
\rho_{12}^*(t)&\rho_{22}(t)
\end{matrix}\right),
\] 
given by the Lindblad equation
\beq\label{m1}
\frac{d}{dt}\rho(t)=\mathcal{L}_{dep}(\rho(t))=\sum_{i=1}^3\gamma_i(t)\left[\sigma_i\rho(t)\sigma_i-\rho(t)\right],
\eeq
where $\gamma_i(t)$'s are the time-dependent Lindblad coefficients and $\sigma_i$'s are the usual Pauli matrices. Further, let us look into a special case with $\gamma_1(t)=\gamma_2(t)=\Gamma_1(t)/2$ and $\gamma_3(t)=-\Gamma_2(t)/2$, where $\Gamma_1(t),~\Gamma_2(t)$ are  always positive. Solving this master equation, we come up with the following completely positive trace-preserving dynamical map,
\begin{align}
\rho_{11}(t)=&\rho_{11}(0)\left(\frac{1+e^{-2\zeta_1 (t)}}{2}\right)+\rho_{22}(0)\left(\frac{1-e^{-2\zeta_1(t)}}{2}\right),\nonumber\\
\rho_{22}(t)=&1-\rho_{11}(t),\quad\rho_{12}(t)=\rho_{12}(0)e^{-\zeta_2(t)},\label{m2}
\end{align}
 where $\zeta_1(t)=\int_0^t\Gamma_1(s)ds$ and $\zeta_2(t)=\int_0^t(\Gamma_1(s)-\Gamma_2(s))ds$, respectively. Using Choi isomorphism~\citep{choi75,jamil72}, one can verify that this map is completely positive for some choices of $\Gamma_1(t)$ and $\Gamma_2(t)$. For example, one such choice would be $\Gamma_1(t)=1$ and $\Gamma_2(t)=\tanh(t)$~\citep{hall14}.
  Hence, for such choices, the map represents a legitimate quantum channel. If we consider the dynamical map for an intermediate time gap $t\rightarrow t+\epsilon$, where $\epsilon$ is small such that $\Gamma_i\epsilon\ll 1$, the dynamical map can be expressed as 
\beq\label{m3}
\rho(t+\epsilon)\approx [\mathcal{I}+\epsilon\mathcal{L}_{dep}]\left(\rho(t)\right)
\eeq
by neglecting the higher-order terms of the expansion. Here $\mathcal{I}(\cdot)$ denotes the identity operation. If we check the eigenvalues of the Choi state~\citep{choi75,jamil72} for the intermediate map, we will find that one of its eigenvalues is negative and amounts to $-\epsilon\Gamma_2(t)$. Therefore, it is clear that the intermediate map is not completely positive. For this map to be positive, the matrix $\rho(t+\epsilon)$ has to be positive for any arbitrary input $\rho(t)$, which presents us with the following inequality,
\[
1-\left(1-2(\alpha-\beta)\right)^2\geq 0,
\]
where we have used the shorthand $\alpha\equiv\epsilon\Gamma_1(t)/2$ and  $\beta\equiv\epsilon\Gamma_2(t)/2$. Note that $\alpha$ and $\beta$ are time-dependent quantities, but 
we drop their time functional representation for brevity. The condition $\Gamma_1(t)\geq\Gamma_2(t)$ implies $\beta \leq \alpha \ll1$.  Upon simplifying, we find the following condition for positivity:
\beq\label{m4}
|\alpha-\beta|=\frac{\epsilon}2|\Gamma_1(t)-\Gamma_2(t)|\leq \frac{1}{2}.
\eeq
It is thus evident that this map can detect bipartite entanglement. Below, we demonstrate a simple example to show the efficiency of this map for entanglement detection. 

We can consider the two qubit Werner state, $\rho_w=p\frac{\mathbb{I}}4+(1-p)\ket{\psi}\bra{\psi} $ where $p$ denotes the probability of mixing, $\mathbb{I}$ stands for the $4$-dimensional identity matrix, and $\ket{\psi}=(\ket{00}+\ket{11})/\sqrt{2}$ is one of the maximally entangled two-qubit states. The condition to identify this state as entangled can be expressed as
\beq\label{m5}
 \beta >\frac{p}{4(1-p)}.
\eeq

We now move on to the main subject of this paper, i.e., the detection of GME.

\section{Detection of genuine multipartite entanglement} \label{sec:detection}
\noindent
Let us first appreciate the complexity of multipartite entanglement detection. The strongest notion of separability in a multipartite quantum state is ``full separability''. A $d$-partite state is fully separable if (and only if) it can be written as
\[\rho_{sep}=\sum_i p_i~\rho_1^i\otimes\rho_2^i\cdots\otimes\rho_d^i,\]
with $\{p_i\}$ being the probability distribution. If a quantum state cannot be expressed in the form of $\rho_{sep}$, then the state is in some way entangled. However, that does not necessarily imply that the state is genuinely entangled. As an example, we can consider a bi-separable state $\rho_{2\mbox{-}sep}$, which can be expressed as
\[ \rho_{2\mbox{-}sep}=\sum_i\mu_i\rho_1^i\otimes\sigma^i, \]
where $\{\mu_i\}$ is the probability distribution, and $\sigma^i$'s are the states containing every other partition except the first, which can very well be entangled. However, despite being entangled, this state is not genuinely entangled. A $d$-partite entangled state, which cannot be expressed as a bi-separable state, is called a \emph{genuinely} multipartite entangled state. Interestingly, since the set of bi-separable quantum states form a convex closed set, by the Hann-Banach theorem, linear Hermitian witnesses can be constructed to detect GME. However, positive maps fail to detect GME for a very simple reason---if a positive map is applied on one of the partitions of a multipartite quantum system, the output matrix can also acquire negative eigenvalues even if the state is bi-separable. Thus, positive maps cannot distinguish between bi-separable and genuinely multipartite entangled states. 

To surpass this problem, we employ a method \citep{sengupta14,huber17} that eliminates the cases of bi-separable states and hence enables us to single out the GME states with a \emph{sufficient} condition, similar to the one given by positive maps in the case of bipartite entanglement.
The main entanglement criterion based on  positive maps is that if we extend them  to the bipartite system $\Phi:=\Lambda_A\otimes \mathcal{I}_B$, they remain positive for the separable states, but become non-positive for some entangled states. 

A map may detect GME if the following condition is satisfied,
\[\Phi[\rho_{2\mbox{-}sep}]\geq0\quad\forall \rho_{2\mbox{-}sep}.\]
From~\citep{sengupta14}, we know that for detecting $\rho_{GME}$, it suffices to find a GME-witness of the form:
\[ W_{GME} = \sum_A \left(\Lambda_A^*\otimes \mathcal{I}_{\bar A}\right)(\ket{\psi_A}\bra{\psi_A})+M_{\{\Lambda_A,\ket{\psi_A}\}_A}\]
where for each partition $A,\ \Lambda_A$ is a positive map and $\ket{\psi_A}$ is chosen such that $\bra{\psi_A}\Lambda_A\otimes \mathcal{I}_{\bar A}(\rho_{GME})\ket{\psi_A}< 0$, $M_{\{\Lambda_A,\ket{\psi_A}\}_A}$ is a positive operator depending on the choice of $\Lambda_A$ and $\ket{\psi_A}$. Here $\mathcal{I}_{\bar A}(\cdot)$ denotes the identity operation on the complimentary partition $\bar A$. One can find how $W_{GME}$ is obtained and, in particular, $M_{\{\Lambda_A,\ket{\psi_A}\}_A}$ is constructed from~\citep{huber17}.
Motivated by this, we seek for a similar map:
\begin{align}
\Phi_{GME} = \sum_{A} \Lambda_A\otimes \mathcal{I}_{\bar A}\circ \mathcal{U}^{(A)}\otimes \mathcal I_{\bar A} + M  
\end{align} 
where $M$ is a positive map, $\mathcal{U}^{(A)}[\rho] := \sum_i p_i^{(A)}U_i^{(A)}\rho (U_i^{(A)})^\dagger$ is a family of convex combinations of local unitaries, and $\Phi_{GME}[\rho_{2\mbox{-}sep}] \geq 0 $ $\forall \rho_{2\mbox{-}sep}$.

\subsection{A GME-detecting map}
\noindent Here, based on the eternal non-Markovian evolution described in the previous section, we construct a GME-detecting map.
Let  $\Lambda[\rho] = \rho + \alpha(\sigma_x\rho\sigma_x - \rho + \sigma_y\rho\sigma_y - \rho) - \beta(\sigma_z\rho\sigma_z - \rho)$, i.e.,
for $\rho = \begin{pmatrix} \rho_{11}&\rho_{12}\\\rho_{21}&\rho_{22}\\  \end{pmatrix}$, we define
\begin{align}\hspace{-0.15cm}\Lambda[\rho] =&  \begin{pmatrix}
(1-2\alpha)\rho_{11} + 2\alpha\rho_{22} &(1-2\alpha+2\beta)\rho_{12} \\ (1-2\alpha+2\beta)\rho_{21} & 2\alpha\rho_{11} + (1-2\alpha)\rho_{22}
\end{pmatrix}.
\end{align}
We can construct a map using $\Lambda$, $M = c \cdot \Tr$, and setting $\mathcal{U}^{(A)}$ to be an identity operation in a $3$-qubit setting:
\begin{align}
\Phi_{\Lambda} =&\ \Lambda_A\otimes \mathcal{I}_B\otimes \mathcal{I}_C + \mathcal{I}_A \otimes \Lambda_B\otimes \mathcal{I}_C + \mathcal{I}_A\otimes \mathcal{I}_B\otimes\Lambda_C\nonumber\\&
+c\cdot \mathbb{I}\cdot\Tr 
\label{eq:mapc}
\end{align}
where $c$ is a constant and the  subscripts $A$, $B$ and $C$ stand for the  three parties.
We choose the constant $c$ such that the map $\Phi_{\Lambda}$ gives positive outcome for all biseparable states. To figure out a suitable value of $c$, we make the following two observations:\medskip

\noindent
{\bf Statement~$1$:} The minimum eigenvalue of
$\Lambda_A\otimes \mathcal{I}_{\bar A}$ acting on any two qubit state
is $-\beta$ for maximally entangled state.

\proof To prove this statement, we need to borrow the properties of the minimal output eigenvalues of positive maps \citep{huber17}. Let us consider $\nu(\chi)$ to be the minimum output eigenvalue for a positive map $\chi$, i.e.,  
\[\nu(\chi)=\min_{\sigma_{AB}}\,EV_{min}\Big(\left(\chi_A\otimes\mathcal{I}_B\right)\left[\sigma_{AB}\right]\Big),\]
where $A$ and $B$ are the two parties of an arbitrary bipartite quantum system and $EV_{min}$ denotes the minimum eigenvalue. It is also known that $\nu(\chi)$ can be determined by minimizing only over the pure states \citep{huber17}. Thus, for the qubit map $\Lambda$, $\nu(\Lambda)$ is obtained by minimizing over two-qubit pure states. But for pure product states, the minimum eigenvalue is always non-negative since $\Lambda$ is a positive map. Therefore, it is sufficient to consider only two-qubit pure entangled states. Such an entangled state can be parametrized as 
\[ \ket\psi_e = \sum_{i=0}^1 c_i \ket i\ket i,~~\mbox{with}~~ |c_0|^2+|c_1|^2=1.\]
Applying the map $\Lambda_A\otimes \mathcal I_B$ on  $\ket\psi_e$, we get the smaller eigenvalue as
\begin{align*}
   \frac{1}{2}\Bigg[(1-2\alpha)-\sqrt{(1-2\alpha)^2+16|c_0|^2|c_1|^2\beta(1-2\alpha+\beta)}\Bigg].
\end{align*}
Clearly, this is minimum when $|c_0|^2|c_1|^2$ is maximum, i.e., $|c_0|=|c_1|=1/\sqrt{2}$. Therefore, without any loss of generality, we can omit the phase factor and take $c_0=c_1=1/\sqrt{2}$, which gives us the maximally entangled state. This final matrix obtain upon the application of the positive map, also happens to be the Choi state for the same map~\citep{choi75,jamil72}. Plugging this value we get the minimum eigenvalue, $\nu(\Lambda)=-\beta$. \qed \\

\noindent
{\bf Statement~$2$:} For all bi-separable $3$-qubit states $\rho_{2\mbox{-}sep}$, the minimum value of $c$ for which $\Phi_{\Lambda}[\rho_{2\mbox{-}sep}] \geq 0 $ is $ 2\beta$. Here the map $\Phi_{\Lambda}$ is defined in equation \eqref{eq:mapc}.

\proof 
To prove the second statement, let us assume $\rho_{2\mbox{-}sep}$ is of the form, 
\begin{align*}
\rho_{2\mbox{-}sep} = \sum_{\kappa=1}^n\sum_{X\in\{A,B,C\}}p_X^{\kappa}\cdot\rho_X^{\kappa}\otimes\rho_{\bar X}^{\kappa}  
\end{align*}
where $\{p^\kappa_X\}$ is the probability distribution, $\rho_X^{\kappa}$ is the qubit density matrix of the party $X$ and $\rho_{\bar X}^{\kappa}$ is the corresponding complementary two qubit quantum state. The sum $\sum_X$ runs  over all biseparations $X|\bar X$ for $X\in\{A,B,C\}$. 
Since $\rho_{2\mbox{-}sep}$ is a density matrix,  $\Tr[\rho_{2\mbox{-}sep}] =1$. Now, we can write
\begin{align}
 \Phi_{\Lambda}[\rho_{2\mbox{-}sep}] =\hspace{-0.45cm} \sum_{J\in \{A,B,C\}} \left(\Lambda_J \otimes \mathcal{I}_{\bar J}\right)(\rho_{2\mbox{-}sep}) 
  + c\cdot\mathbb{I}\cdot \Tr[\rho_{2\mbox{-}sep}].
\end{align}
This gives us,
\begin{align}
&\hspace{-0.75cm}EV_{min}\Big(\Phi_{\Lambda}[\rho_{2\mbox{-}sep}]\Big) 
\nonumber\\
\geq&\sum_{\kappa, J} p_J^{\kappa}\, \overbrace{  EV_{min}\Big(\left(\Lambda_J \otimes \mathcal{I}_{\bar J}\right)[\rho_J^{\kappa}\otimes\rho_{\bar J}^{\kappa}]\Big)}^{\geq0}\nonumber\\
&+\overbrace{
\sum_{\substack{\kappa, J,X\\ J\neq X}} p_X^{\kappa}\, EV_{min}\Big( \left(\Lambda_J\otimes \mathcal{I}_{\bar J}\right)[\rho_X^{\kappa}\otimes\rho_{\bar X}^{\kappa}] \Big)}^{\geq -2\beta }+ c, \label{eqNN}
\end{align}
where we have used Statement 1, which states the fact that for an arbitrary two qubit state $\rho_{AB}$ and the corresponding Choi state $\rho_{C}$,
\begin{align}
  EV_{min}\Big(\left(\Lambda_A\otimes\mathcal I_B\right)\left[\rho_{AB}\right]\Big) \geq EV_{min}\Big(\left(\Lambda_A\otimes\mathcal I_B\right)\left[\rho_{C}\right]\Big)=-\beta.\nonumber  
\end{align}
Eq.~\eqref{eqNN} tells us that for an arbitrary biseparable state, a negative contribution to the minimum eigenvalue might come from the entangled partition, which needs to be compensated by a suitable positive choice of $c$. Clearly, if $c\geq2\beta$ then $
EV_{min}\Big(\Phi_{\Lambda}[\rho_{2\mbox{-}sep}]\Big)\geq 0$. In other words,  $c=2\beta$ is the minimal choice.\qed\\
\medskip

\begin{figure}[t]
\centering\includegraphics[width=\columnwidth]{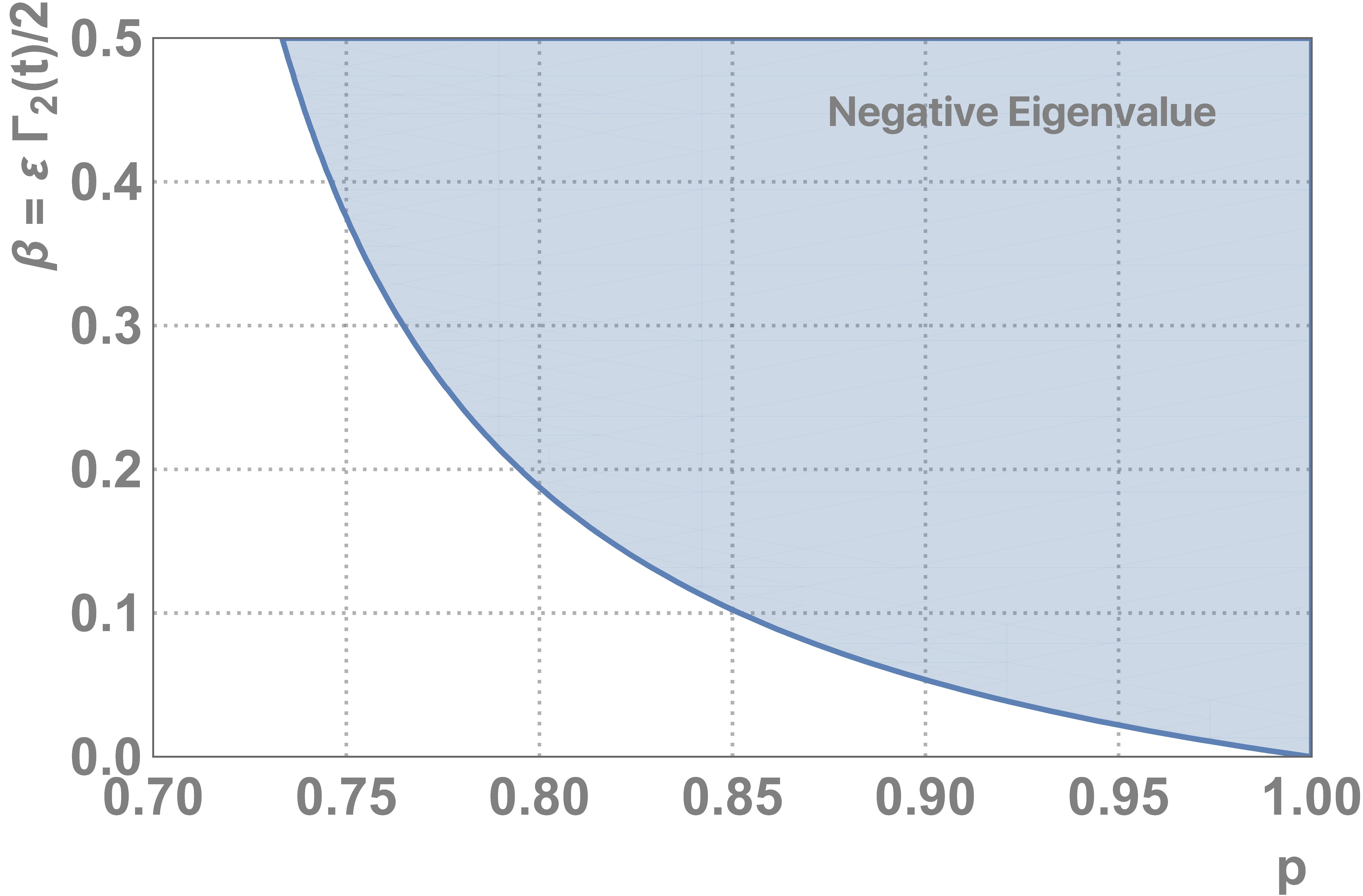}
\caption{Application of the map $\Phi_\Lambda$ on the mixed density matrix $\rho^p_\gamma$ [Eq.~\eqref{eq:evGHZpden}] makes the smallest eigenvalue [Eq.~\eqref{eq:evGHZp}] negative in the shaded region. The parameter $\beta$ is related to the Lindblad coefficients as $\beta = \epsilon\Gamma_2(t)/2$ with $\epsilon$ being a small time step. 
}\label{fig:nEV}
\end{figure}


\begin{figure}
\centering
\includegraphics[width=\columnwidth]{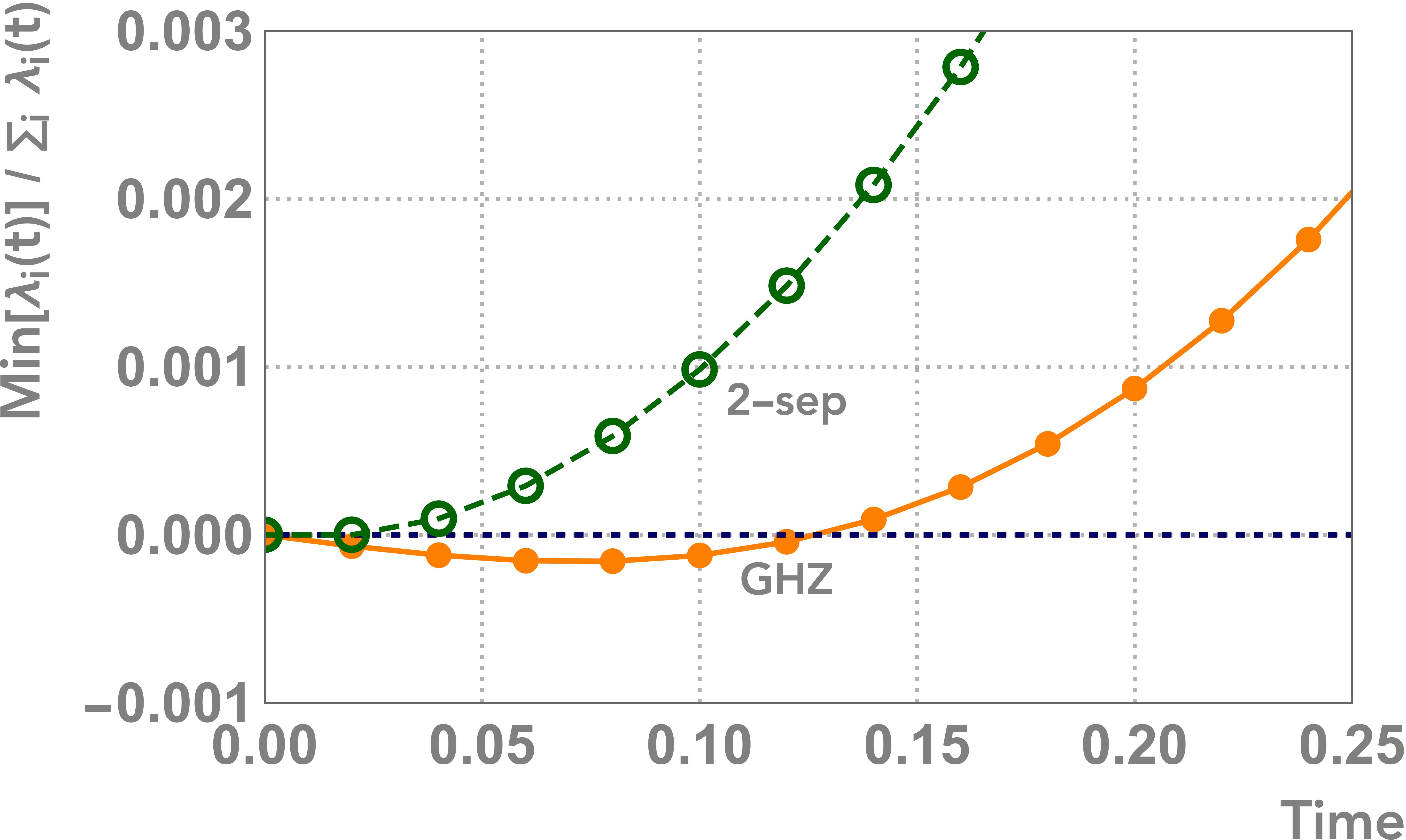}
\caption{We show the time evolution of the minimum eigenvalues for a GHZ and a bi-separable state under the map $\Phi_\Lambda$ [Eq.~\eqref{eq:map}] numerically for a step size $\epsilon =1/50$ time unit with $\Gamma_1(t)=1$ and $\Gamma_2(t)=\tanh(t)$. When we start from the GHZ state (solid line), the minimum eigenvalue initially turns negative. 
}\label{fig:minEV}
\end{figure}
The threshold choice of $c$ ensures that the map $\Phi_{\Lambda}$ gives positive outcome for all $3$-qubit biseparable states. Therefore, we set $c=2\beta$ in Eq.~\eqref{eq:mapc} to obtain the final map,
\begin{align}
\Phi_{\Lambda} =&\ \Lambda_A\otimes \mathcal{I}_B\otimes \mathcal{I}_C + \mathcal{I}_A \otimes \Lambda_B\otimes \mathcal{I}_C + \mathcal{I}_A\otimes \mathcal{I}_B\otimes\Lambda_C\nonumber\\&+2\beta\cdot\mathbb{I}\cdot \Tr.
\label{eq:map}
\end{align}
This map can detect GME; for example, it can do so from the GHZ state,
\begin{align}
\ket{\gamma} =& \frac{1}{\sqrt{2}}\Big(\ket{000} + \ket{111}\Big),\ \rho_\gamma = \ket{\gamma}\bra{\gamma}.\label{eq:GHZ} 
\end{align}
One can check that
\begin{align}
EV_{min}\Big(\Phi_{\Lambda}\left[\rho_\gamma\right]\Big) =-\beta.\label{eq:evGHZ}
\end{align}
The map is robust in the sense that it can also detect GME from a noisy GHZ state, 
\begin{align}\label{eq:evGHZpden}
    \rho_\gamma^p  =p \ket{\gamma}\bra{\gamma} + \Big(1-p\Big)\frac{\mathbb{I}_8}{8}.
\end{align} 
In this case, the minimum eigenvalue generalizes to
\begin{align}\label{eq:evGHZp}
EV_{min}\Big(\Phi_{\Lambda}[\rho_\gamma^p]\Big) =\frac{1}{8}\left[3(1-p)+(2-3p)\beta\right].
\end{align}
As the minimum eigenvalue is independent of $\alpha$ (except for the condition that $\alpha\geq \beta$), we show the region where the map is useful, i.e., it has a negative eigenvalue in the $p$-$\beta$ plane in Fig.~\ref{fig:nEV}. In Fig.~\ref{fig:minEV}, we plot the time evolution of the minimum eigenvalues of a  GHZ and a bi-separable state under $\Phi_\Lambda$ [Eq.~\eqref{eq:map}] numerically for a step size $\epsilon =1/50$ time unit. We see that the minimum eigenvalue initially turns negative for the GHZ state.  

\subsection{GME detection through a witness operator}
\noindent
One of the most prominent ways to detect entanglement experimentally is via the application of witness operators \citep{wit1,wit2,wit3,wit4,wit5,wit6,guhne09,wit8,wit9,wit10}. The Hann-Banach theorem tells us that elements outside a convex and compact set can always be separated by a hyperplane, which can be realized physically by linear hermitian operators. Similar to the case of bipartite states,  all bi-separable states form a convex and compact set in the multipartite scenario as well. Hence, the GME states can be detected by linear hermitian witness operators. Below, we construct a GME witness operator from the non-Markovianity induced GME detection map $\Phi_{\Lambda}$.\\

\noindent\textbf{Statement~$3$:} The harmitian operator 
$\Phi_{\Lambda}(\ket{\tilde{\gamma}}\bra{\tilde{\gamma}})$, with $\ket{\tilde{\gamma}}=\frac{1}{\sqrt{2}}(\ket{000}-\ket{111})$, is a GME witness.

\proof Let $\sigma_{2-sep}$ be a $3$-qubit bi-separable state. Therefore, $\Phi_{\Lambda}(\sigma_{2-sep})$ is a positive semi-definite matrix with 
\[
\Tr \left[\Phi_{\Lambda}(\sigma_{2-sep})\ket{\tilde{\gamma}}\bra{\tilde{\gamma}}\right]\geq 0.
\]
Using the Lindblad structure of the $\Lambda_A,~\Lambda_B,~\Lambda_C$ maps from Eq.~\eqref{eq:map}, it can be shown that 
\[\Tr \left[\Phi_{\Lambda}(\sigma_{2-sep})\ket{\tilde{\gamma}}\bra{\tilde{\gamma}}\right]= \Tr \left[\Phi_{\Lambda}(\ket{\tilde{\gamma}}\bra{\tilde{\gamma}})\sigma_{2-sep}\right]\geq 0.\]
The result holds for any arbitrary $3$-qubit bi-separable state. If we now apply the harmitian operator $\mathcal{W}_{GME}=\Phi_{\Lambda}(\ket{\tilde{\gamma}}\bra{\tilde{\gamma}})$ on the state $\rho_\gamma^p$ defined in Eq.~\eqref{eq:evGHZpden}, we get 
\[\Tr \left[\mathcal{W}_{GME}\,\rho_\gamma^p\right]=\frac{1}{8}\left[3(1-p)+(2-3p)\beta\right],\]
which, for the GHZ state ($p=1$), amounts to $-\beta$. Therefore, the witness operator $\mathcal{W}_{GME}$ succesfully detects the GHZ state. In fact, it is evident from Eq.~\eqref{eq:evGHZp} that the witness detects the full range of states detected by the map $\Phi_{\Lambda}$, which proves $\mathcal{W}_{GME}$ to be a GME witness.\qed \\

Since the GME witness operator proposed in \textbf{Statement~$3$} can be constructed experimentally,  our method of GME detection through non-Markovianity can be implemented in a practical scenario.

\section{Summary and conclusions}
\noindent
In this study, we have constructed a novel method to detect genuine multipartite entanglement in $3$-qubit systems by harnessing eternal non-Markovian operations. We have used the properties of $p$-divisible dynamics and the theory of entanglement detection by positive maps to construct our method. We have further devised a witness operator from this map. In this paper, we have only considered $3$-qubit entangled states, but our method can be extended to more complex higher-dimensional multipartite entangled states as well. 
To do so, one has to first construct a positive map out of some higher-dimensional eternal non-Markovian operation that will give a lower bound on the minimum negative eigenvalue when applied to the system. It is then straightforward to adapt our method for the corresponding higher-dimensional system. Our study provides just a glimpse of the deep connection between quantum non-Markovianity and entanglement. It opens up the possibility of further exploring the relationship between the two fundamental quantum information-theoretic domains of study.  

\acknowledgements
\noindent
We thank Professor M. J. W. Hall for his helpful comments.


\bibliographystyle{apsrev4-1}
\bibliography{sisterP}

\begin{thebibliography}{55}%
\makeatletter
\providecommand \@ifxundefined [1]{%
 \@ifx{#1\undefined}
}%
\providecommand \@ifnum [1]{%
 \ifnum #1\expandafter \@firstoftwo
 \else \expandafter \@secondoftwo
 \fi
}%
\providecommand \@ifx [1]{%
 \ifx #1\expandafter \@firstoftwo
 \else \expandafter \@secondoftwo
 \fi
}%
\providecommand \natexlab [1]{#1}%
\providecommand \enquote  [1]{``#1''}%
\providecommand \bibnamefont  [1]{#1}%
\providecommand \bibfnamefont [1]{#1}%
\providecommand \citenamefont [1]{#1}%
\providecommand \href@noop [0]{\@secondoftwo}%
\providecommand \href [0]{\begingroup \@sanitize@url \@href}%
\providecommand \@href[1]{\@@startlink{#1}\@@href}%
\providecommand \@@href[1]{\endgroup#1\@@endlink}%
\providecommand \@sanitize@url [0]{\catcode `\\12\catcode `\$12\catcode
  `\&12\catcode `\#12\catcode `\^12\catcode `\_12\catcode `\%12\relax}%
\providecommand \@@startlink[1]{}%
\providecommand \@@endlink[0]{}%
\providecommand \url  [0]{\begingroup\@sanitize@url \@url }%
\providecommand \@url [1]{\endgroup\@href {#1}{\urlprefix }}%
\providecommand \urlprefix  [0]{URL }%
\providecommand \Eprint [0]{\href }%
\providecommand \doibase [0]{http://dx.doi.org/}%
\providecommand \selectlanguage [0]{\@gobble}%
\providecommand \bibinfo  [0]{\@secondoftwo}%
\providecommand \bibfield  [0]{\@secondoftwo}%
\providecommand \translation [1]{[#1]}%
\providecommand \BibitemOpen [0]{}%
\providecommand \bibitemStop [0]{}%
\providecommand \bibitemNoStop [0]{.\EOS\space}%
\providecommand \EOS [0]{\spacefactor3000\relax}%
\providecommand \BibitemShut  [1]{\csname bibitem#1\endcsname}%
\let\auto@bib@innerbib\@empty
\bibitem [{\citenamefont {Ekert}(1991)}]{ekert91}%
  \BibitemOpen
  \bibfield  {author} {\bibinfo {author} {\bibfnamefont {A.~K.}\ \bibnamefont
  {Ekert}},\ }\href {\doibase 10.1103/PhysRevLett.67.661} {\bibfield  {journal}
  {\bibinfo  {journal} {Phys. Rev. Lett.}\ }\textbf {\bibinfo {volume} {67}},\
  \bibinfo {pages} {661} (\bibinfo {year} {1991})}\BibitemShut {NoStop}%
\bibitem [{\citenamefont {Bennett}\ and\ \citenamefont
  {Wiesner}(1992)}]{bennett92}%
  \BibitemOpen
  \bibfield  {author} {\bibinfo {author} {\bibfnamefont {C.~H.}\ \bibnamefont
  {Bennett}}\ and\ \bibinfo {author} {\bibfnamefont {S.~J.}\ \bibnamefont
  {Wiesner}},\ }\href {\doibase 10.1103/PhysRevLett.69.2881} {\bibfield
  {journal} {\bibinfo  {journal} {Phys. Rev. Lett.}\ }\textbf {\bibinfo
  {volume} {69}},\ \bibinfo {pages} {2881} (\bibinfo {year}
  {1992})}\BibitemShut {NoStop}%
\bibitem [{\citenamefont {Bennett}\ \emph {et~al.}(1993)\citenamefont
  {Bennett}, \citenamefont {Brassard}, \citenamefont {Cr\'epeau}, \citenamefont
  {Jozsa}, \citenamefont {Peres},\ and\ \citenamefont {Wootters}}]{bennett93}%
  \BibitemOpen
  \bibfield  {author} {\bibinfo {author} {\bibfnamefont {C.~H.}\ \bibnamefont
  {Bennett}}, \bibinfo {author} {\bibfnamefont {G.}~\bibnamefont {Brassard}},
  \bibinfo {author} {\bibfnamefont {C.}~\bibnamefont {Cr\'epeau}}, \bibinfo
  {author} {\bibfnamefont {R.}~\bibnamefont {Jozsa}}, \bibinfo {author}
  {\bibfnamefont {A.}~\bibnamefont {Peres}}, \ and\ \bibinfo {author}
  {\bibfnamefont {W.~K.}\ \bibnamefont {Wootters}},\ }\href {\doibase
  10.1103/PhysRevLett.70.1895} {\bibfield  {journal} {\bibinfo  {journal}
  {Phys. Rev. Lett.}\ }\textbf {\bibinfo {volume} {70}},\ \bibinfo {pages}
  {1895} (\bibinfo {year} {1993})}\BibitemShut {NoStop}%
\bibitem [{\citenamefont {Walter}\ \emph {et~al.}(2017)\citenamefont {Walter},
  \citenamefont {Gross},\ and\ \citenamefont {Eisert}}]{multi1}%
  \BibitemOpen
  \bibfield  {author} {\bibinfo {author} {\bibfnamefont {M.}~\bibnamefont
  {Walter}}, \bibinfo {author} {\bibfnamefont {D.}~\bibnamefont {Gross}}, \
  and\ \bibinfo {author} {\bibfnamefont {J.}~\bibnamefont {Eisert}},\
  }\href@noop {} {\enquote {\bibinfo {title} {Multi-partite entanglement},}\ }
  (\bibinfo {year} {2017}),\ \Eprint {http://arxiv.org/abs/1612.02437}
  {arXiv:1612.02437 [quant-ph]} \BibitemShut {NoStop}%
\bibitem [{\citenamefont {Enr{\'{\i}}quez}\ \emph {et~al.}(2016)\citenamefont
  {Enr{\'{\i}}quez}, \citenamefont {Wintrowicz},\ and\ \citenamefont
  {{\.{Z}}yczkowski}}]{multi2}%
  \BibitemOpen
  \bibfield  {author} {\bibinfo {author} {\bibfnamefont {M.}~\bibnamefont
  {Enr{\'{\i}}quez}}, \bibinfo {author} {\bibfnamefont {I.}~\bibnamefont
  {Wintrowicz}}, \ and\ \bibinfo {author} {\bibfnamefont {K.}~\bibnamefont
  {{\.{Z}}yczkowski}},\ }\href {\doibase 10.1088/1742-6596/698/1/012003}
  {\bibfield  {journal} {\bibinfo  {journal} {Journal of Physics: Conference
  Series}\ }\textbf {\bibinfo {volume} {698}},\ \bibinfo {pages} {012003}
  (\bibinfo {year} {2016})}\BibitemShut {NoStop}%
\bibitem [{\citenamefont {Arnaud}\ and\ \citenamefont {Cerf}(2013)}]{multi3}%
  \BibitemOpen
  \bibfield  {author} {\bibinfo {author} {\bibfnamefont {L.}~\bibnamefont
  {Arnaud}}\ and\ \bibinfo {author} {\bibfnamefont {N.~J.}\ \bibnamefont
  {Cerf}},\ }\href {\doibase 10.1103/PhysRevA.87.012319} {\bibfield  {journal}
  {\bibinfo  {journal} {Phys. Rev. A}\ }\textbf {\bibinfo {volume} {87}},\
  \bibinfo {pages} {012319} (\bibinfo {year} {2013})}\BibitemShut {NoStop}%
\bibitem [{\citenamefont {Amico}\ \emph {et~al.}(2008)\citenamefont {Amico},
  \citenamefont {Fazio}, \citenamefont {Osterloh},\ and\ \citenamefont
  {Vedral}}]{multi5}%
  \BibitemOpen
  \bibfield  {author} {\bibinfo {author} {\bibfnamefont {L.}~\bibnamefont
  {Amico}}, \bibinfo {author} {\bibfnamefont {R.}~\bibnamefont {Fazio}},
  \bibinfo {author} {\bibfnamefont {A.}~\bibnamefont {Osterloh}}, \ and\
  \bibinfo {author} {\bibfnamefont {V.}~\bibnamefont {Vedral}},\ }\href
  {\doibase 10.1103/RevModPhys.80.517} {\bibfield  {journal} {\bibinfo
  {journal} {Rev. Mod. Phys.}\ }\textbf {\bibinfo {volume} {80}},\ \bibinfo
  {pages} {517} (\bibinfo {year} {2008})}\BibitemShut {NoStop}%
\bibitem [{\citenamefont {Goyeneche}\ \emph {et~al.}(2015)\citenamefont
  {Goyeneche}, \citenamefont {Alsina}, \citenamefont {Latorre}, \citenamefont
  {Riera},\ and\ \citenamefont {\ifmmode~\dot{Z}\else
  \.{Z}\fi{}yczkowski}}]{multi6}%
  \BibitemOpen
  \bibfield  {author} {\bibinfo {author} {\bibfnamefont {D.}~\bibnamefont
  {Goyeneche}}, \bibinfo {author} {\bibfnamefont {D.}~\bibnamefont {Alsina}},
  \bibinfo {author} {\bibfnamefont {J.~I.}\ \bibnamefont {Latorre}}, \bibinfo
  {author} {\bibfnamefont {A.}~\bibnamefont {Riera}}, \ and\ \bibinfo {author}
  {\bibfnamefont {K.}~\bibnamefont {\ifmmode~\dot{Z}\else
  \.{Z}\fi{}yczkowski}},\ }\href {\doibase 10.1103/PhysRevA.92.032316}
  {\bibfield  {journal} {\bibinfo  {journal} {Phys. Rev. A}\ }\textbf {\bibinfo
  {volume} {92}},\ \bibinfo {pages} {032316} (\bibinfo {year}
  {2015})}\BibitemShut {NoStop}%
\bibitem [{\citenamefont {Goyeneche}\ and\ \citenamefont {\ifmmode~\dot{Z}\else
  \.{Z}\fi{}yczkowski}(2014)}]{multi7}%
  \BibitemOpen
  \bibfield  {author} {\bibinfo {author} {\bibfnamefont {D.}~\bibnamefont
  {Goyeneche}}\ and\ \bibinfo {author} {\bibfnamefont {K.}~\bibnamefont
  {\ifmmode~\dot{Z}\else \.{Z}\fi{}yczkowski}},\ }\href {\doibase
  10.1103/PhysRevA.90.022316} {\bibfield  {journal} {\bibinfo  {journal} {Phys.
  Rev. A}\ }\textbf {\bibinfo {volume} {90}},\ \bibinfo {pages} {022316}
  (\bibinfo {year} {2014})}\BibitemShut {NoStop}%
\bibitem [{\citenamefont {Facchi}\ \emph {et~al.}(2008)\citenamefont {Facchi},
  \citenamefont {Florio}, \citenamefont {Parisi},\ and\ \citenamefont
  {Pascazio}}]{multi8}%
  \BibitemOpen
  \bibfield  {author} {\bibinfo {author} {\bibfnamefont {P.}~\bibnamefont
  {Facchi}}, \bibinfo {author} {\bibfnamefont {G.}~\bibnamefont {Florio}},
  \bibinfo {author} {\bibfnamefont {G.}~\bibnamefont {Parisi}}, \ and\ \bibinfo
  {author} {\bibfnamefont {S.}~\bibnamefont {Pascazio}},\ }\href {\doibase
  10.1103/PhysRevA.77.060304} {\bibfield  {journal} {\bibinfo  {journal} {Phys.
  Rev. A}\ }\textbf {\bibinfo {volume} {77}},\ \bibinfo {pages} {060304}
  (\bibinfo {year} {2008})}\BibitemShut {NoStop}%
\bibitem [{\citenamefont {Horodecki}\ \emph {et~al.}(2009)\citenamefont
  {Horodecki}, \citenamefont {Horodecki}, \citenamefont {Horodecki},\ and\
  \citenamefont {Horodecki}}]{wit6}%
  \BibitemOpen
  \bibfield  {author} {\bibinfo {author} {\bibfnamefont {R.}~\bibnamefont
  {Horodecki}}, \bibinfo {author} {\bibfnamefont {P.}~\bibnamefont
  {Horodecki}}, \bibinfo {author} {\bibfnamefont {M.}~\bibnamefont
  {Horodecki}}, \ and\ \bibinfo {author} {\bibfnamefont {K.}~\bibnamefont
  {Horodecki}},\ }\href {\doibase 10.1103/RevModPhys.81.865} {\bibfield
  {journal} {\bibinfo  {journal} {Rev. Mod. Phys.}\ }\textbf {\bibinfo {volume}
  {81}},\ \bibinfo {pages} {865} (\bibinfo {year} {2009})}\BibitemShut
  {NoStop}%
\bibitem [{\citenamefont {Gühne}\ and\ \citenamefont {Tóth}(2009)}]{guhne09}%
  \BibitemOpen
  \bibfield  {author} {\bibinfo {author} {\bibfnamefont {O.}~\bibnamefont
  {Gühne}}\ and\ \bibinfo {author} {\bibfnamefont {G.}~\bibnamefont {Tóth}},\
  }\href {\doibase https://doi.org/10.1016/j.physrep.2009.02.004} {\bibfield
  {journal} {\bibinfo  {journal} {Physics Reports}\ }\textbf {\bibinfo {volume}
  {474}},\ \bibinfo {pages} {1} (\bibinfo {year} {2009})}\BibitemShut {NoStop}%
\bibitem [{\citenamefont {Brukner}\ \emph {et~al.}(2004)\citenamefont
  {Brukner}, \citenamefont {{\. Z}ukowski}, \citenamefont {Pan},\ and\
  \citenamefont {Zeilinger}}]{brukner04}%
  \BibitemOpen
  \bibfield  {author} {\bibinfo {author} {\bibfnamefont {{\v C}.}~\bibnamefont
  {Brukner}}, \bibinfo {author} {\bibfnamefont {M.}~\bibnamefont {{\.
  Z}ukowski}}, \bibinfo {author} {\bibfnamefont {J.-W.}\ \bibnamefont {Pan}}, \
  and\ \bibinfo {author} {\bibfnamefont {A.}~\bibnamefont {Zeilinger}},\ }\href
  {\doibase 10.1103/PhysRevLett.92.127901} {\bibfield  {journal} {\bibinfo
  {journal} {Phys. Rev. Lett.}\ }\textbf {\bibinfo {volume} {92}},\ \bibinfo
  {pages} {127901} (\bibinfo {year} {2004})}\BibitemShut {NoStop}%
\bibitem [{\citenamefont {Gurvits}(2004)}]{gurvits}%
  \BibitemOpen
  \bibfield  {author} {\bibinfo {author} {\bibfnamefont {L.}~\bibnamefont
  {Gurvits}},\ }\href {\doibase https://doi.org/10.1016/j.jcss.2004.06.003}
  {\bibfield  {journal} {\bibinfo  {journal} {Journal of Computer and System
  Sciences}\ }\textbf {\bibinfo {volume} {69}},\ \bibinfo {pages} {448}
  (\bibinfo {year} {2004})},\ \bibinfo {note} {special Issue on STOC
  2003}\BibitemShut {NoStop}%
\bibitem [{\citenamefont {Gharibian}(2010)}]{sevag}%
  \BibitemOpen
  \bibfield  {author} {\bibinfo {author} {\bibfnamefont {S.}~\bibnamefont
  {Gharibian}},\ }\href@noop {} {\bibfield  {journal} {\bibinfo  {journal}
  {Quantum Info. Comput.}\ }\textbf {\bibinfo {volume} {10}},\ \bibinfo {pages}
  {343–360} (\bibinfo {year} {2010})}\BibitemShut {NoStop}%
\bibitem [{\citenamefont {Horodecki}\ \emph {et~al.}(1996)\citenamefont
  {Horodecki}, \citenamefont {Horodecki},\ and\ \citenamefont
  {Horodecki}}]{wit1}%
  \BibitemOpen
  \bibfield  {author} {\bibinfo {author} {\bibfnamefont {M.}~\bibnamefont
  {Horodecki}}, \bibinfo {author} {\bibfnamefont {P.}~\bibnamefont
  {Horodecki}}, \ and\ \bibinfo {author} {\bibfnamefont {R.}~\bibnamefont
  {Horodecki}},\ }\href {\doibase
  https://doi.org/10.1016/S0375-9601(96)00706-2} {\bibfield  {journal}
  {\bibinfo  {journal} {Physics Letters A}\ }\textbf {\bibinfo {volume}
  {223}},\ \bibinfo {pages} {1 } (\bibinfo {year} {1996})}\BibitemShut
  {NoStop}%
\bibitem [{\citenamefont {Bourennane}\ \emph {et~al.}(2004)\citenamefont
  {Bourennane}, \citenamefont {Eibl}, \citenamefont {Kurtsiefer}, \citenamefont
  {Gaertner}, \citenamefont {Weinfurter}, \citenamefont {G\"uhne},
  \citenamefont {Hyllus}, \citenamefont {Bru\ss{}}, \citenamefont
  {Lewenstein},\ and\ \citenamefont {Sanpera}}]{wit2}%
  \BibitemOpen
  \bibfield  {author} {\bibinfo {author} {\bibfnamefont {M.}~\bibnamefont
  {Bourennane}}, \bibinfo {author} {\bibfnamefont {M.}~\bibnamefont {Eibl}},
  \bibinfo {author} {\bibfnamefont {C.}~\bibnamefont {Kurtsiefer}}, \bibinfo
  {author} {\bibfnamefont {S.}~\bibnamefont {Gaertner}}, \bibinfo {author}
  {\bibfnamefont {H.}~\bibnamefont {Weinfurter}}, \bibinfo {author}
  {\bibfnamefont {O.}~\bibnamefont {G\"uhne}}, \bibinfo {author} {\bibfnamefont
  {P.}~\bibnamefont {Hyllus}}, \bibinfo {author} {\bibfnamefont
  {D.}~\bibnamefont {Bru\ss{}}}, \bibinfo {author} {\bibfnamefont
  {M.}~\bibnamefont {Lewenstein}}, \ and\ \bibinfo {author} {\bibfnamefont
  {A.}~\bibnamefont {Sanpera}},\ }\href {\doibase
  10.1103/PhysRevLett.92.087902} {\bibfield  {journal} {\bibinfo  {journal}
  {Phys. Rev. Lett.}\ }\textbf {\bibinfo {volume} {92}},\ \bibinfo {pages}
  {087902} (\bibinfo {year} {2004})}\BibitemShut {NoStop}%
\bibitem [{\citenamefont {Terhal}(2000)}]{wit3}%
  \BibitemOpen
  \bibfield  {author} {\bibinfo {author} {\bibfnamefont {B.~M.}\ \bibnamefont
  {Terhal}},\ }\href {\doibase https://doi.org/10.1016/S0375-9601(00)00401-1}
  {\bibfield  {journal} {\bibinfo  {journal} {Physics Letters A}\ }\textbf
  {\bibinfo {volume} {271}},\ \bibinfo {pages} {319 } (\bibinfo {year}
  {2000})}\BibitemShut {NoStop}%
\bibitem [{\citenamefont {Lewenstein}\ \emph {et~al.}(2000)\citenamefont
  {Lewenstein}, \citenamefont {Kraus}, \citenamefont {Cirac},\ and\
  \citenamefont {Horodecki}}]{wit4}%
  \BibitemOpen
  \bibfield  {author} {\bibinfo {author} {\bibfnamefont {M.}~\bibnamefont
  {Lewenstein}}, \bibinfo {author} {\bibfnamefont {B.}~\bibnamefont {Kraus}},
  \bibinfo {author} {\bibfnamefont {J.~I.}\ \bibnamefont {Cirac}}, \ and\
  \bibinfo {author} {\bibfnamefont {P.}~\bibnamefont {Horodecki}},\ }\href
  {\doibase 10.1103/PhysRevA.62.052310} {\bibfield  {journal} {\bibinfo
  {journal} {Phys. Rev. A}\ }\textbf {\bibinfo {volume} {62}},\ \bibinfo
  {pages} {052310} (\bibinfo {year} {2000})}\BibitemShut {NoStop}%
\bibitem [{\citenamefont {G\"uhne}\ \emph {et~al.}(2002)\citenamefont
  {G\"uhne}, \citenamefont {Hyllus}, \citenamefont {Bru\ss{}}, \citenamefont
  {Ekert}, \citenamefont {Lewenstein}, \citenamefont {Macchiavello},\ and\
  \citenamefont {Sanpera}}]{wit5}%
  \BibitemOpen
  \bibfield  {author} {\bibinfo {author} {\bibfnamefont {O.}~\bibnamefont
  {G\"uhne}}, \bibinfo {author} {\bibfnamefont {P.}~\bibnamefont {Hyllus}},
  \bibinfo {author} {\bibfnamefont {D.}~\bibnamefont {Bru\ss{}}}, \bibinfo
  {author} {\bibfnamefont {A.}~\bibnamefont {Ekert}}, \bibinfo {author}
  {\bibfnamefont {M.}~\bibnamefont {Lewenstein}}, \bibinfo {author}
  {\bibfnamefont {C.}~\bibnamefont {Macchiavello}}, \ and\ \bibinfo {author}
  {\bibfnamefont {A.}~\bibnamefont {Sanpera}},\ }\href {\doibase
  10.1103/PhysRevA.66.062305} {\bibfield  {journal} {\bibinfo  {journal} {Phys.
  Rev. A}\ }\textbf {\bibinfo {volume} {66}},\ \bibinfo {pages} {062305}
  (\bibinfo {year} {2002})}\BibitemShut {NoStop}%
\bibitem [{\citenamefont {Chru\ifmmode \acute{s}\else
  \'{s}\fi{}ci\ifmmode~\acute{n}\else \'{n}\fi{}ski}\ and\ \citenamefont
  {Sarbicki}(2014)}]{wit8}%
  \BibitemOpen
  \bibfield  {author} {\bibinfo {author} {\bibfnamefont {D.}~\bibnamefont
  {Chru\ifmmode \acute{s}\else \'{s}\fi{}ci\ifmmode~\acute{n}\else
  \'{n}\fi{}ski}}\ and\ \bibinfo {author} {\bibfnamefont {G.}~\bibnamefont
  {Sarbicki}},\ }\href {http://stacks.iop.org/1751-8121/47/i=48/a=483001}
  {\bibfield  {journal} {\bibinfo  {journal} {Journal of Physics A:
  Mathematical and Theoretical}\ }\textbf {\bibinfo {volume} {47}},\ \bibinfo
  {pages} {483001} (\bibinfo {year} {2014})}\BibitemShut {NoStop}%
\bibitem [{\citenamefont {G\"uhne}\ \emph {et~al.}(2007)\citenamefont
  {G\"uhne}, \citenamefont {Lu}, \citenamefont {Gao},\ and\ \citenamefont
  {Pan}}]{wit9}%
  \BibitemOpen
  \bibfield  {author} {\bibinfo {author} {\bibfnamefont {O.}~\bibnamefont
  {G\"uhne}}, \bibinfo {author} {\bibfnamefont {C.-Y.}\ \bibnamefont {Lu}},
  \bibinfo {author} {\bibfnamefont {W.-B.}\ \bibnamefont {Gao}}, \ and\
  \bibinfo {author} {\bibfnamefont {J.-W.}\ \bibnamefont {Pan}},\ }\href
  {\doibase 10.1103/PhysRevA.76.030305} {\bibfield  {journal} {\bibinfo
  {journal} {Phys. Rev. A}\ }\textbf {\bibinfo {volume} {76}},\ \bibinfo
  {pages} {030305} (\bibinfo {year} {2007})}\BibitemShut {NoStop}%
\bibitem [{\citenamefont {Zhang}\ \emph {et~al.}(2008)\citenamefont {Zhang},
  \citenamefont {Zhang}, \citenamefont {Zhang},\ and\ \citenamefont
  {Guo}}]{wit10}%
  \BibitemOpen
  \bibfield  {author} {\bibinfo {author} {\bibfnamefont {C.-J.}\ \bibnamefont
  {Zhang}}, \bibinfo {author} {\bibfnamefont {Y.-S.}\ \bibnamefont {Zhang}},
  \bibinfo {author} {\bibfnamefont {S.}~\bibnamefont {Zhang}}, \ and\ \bibinfo
  {author} {\bibfnamefont {G.-C.}\ \bibnamefont {Guo}},\ }\href {\doibase
  10.1103/PhysRevA.77.060301} {\bibfield  {journal} {\bibinfo  {journal} {Phys.
  Rev. A}\ }\textbf {\bibinfo {volume} {77}},\ \bibinfo {pages} {060301}
  (\bibinfo {year} {2008})}\BibitemShut {NoStop}%
\bibitem [{\citenamefont {Rivas}\ \emph {et~al.}(2010)\citenamefont {Rivas},
  \citenamefont {Huelga},\ and\ \citenamefont {Plenio}}]{rhp1}%
  \BibitemOpen
  \bibfield  {author} {\bibinfo {author} {\bibfnamefont {A.}~\bibnamefont
  {Rivas}}, \bibinfo {author} {\bibfnamefont {S.~F.}\ \bibnamefont {Huelga}}, \
  and\ \bibinfo {author} {\bibfnamefont {M.~B.}\ \bibnamefont {Plenio}},\
  }\href {\doibase 10.1103/PhysRevLett.105.050403} {\bibfield  {journal}
  {\bibinfo  {journal} {Phys. Rev. Lett.}\ }\textbf {\bibinfo {volume} {105}},\
  \bibinfo {pages} {050403} (\bibinfo {year} {2010})}\BibitemShut {NoStop}%
\bibitem [{\citenamefont {Laine}\ \emph {et~al.}(2010)\citenamefont {Laine},
  \citenamefont {Piilo},\ and\ \citenamefont {Breuer}}]{blp1}%
  \BibitemOpen
  \bibfield  {author} {\bibinfo {author} {\bibfnamefont {E.-M.}\ \bibnamefont
  {Laine}}, \bibinfo {author} {\bibfnamefont {J.}~\bibnamefont {Piilo}}, \ and\
  \bibinfo {author} {\bibfnamefont {H.-P.}\ \bibnamefont {Breuer}},\ }\href
  {\doibase 10.1103/PhysRevA.81.062115} {\bibfield  {journal} {\bibinfo
  {journal} {Phys. Rev. A}\ }\textbf {\bibinfo {volume} {81}},\ \bibinfo
  {pages} {062115} (\bibinfo {year} {2010})}\BibitemShut {NoStop}%
\bibitem [{\citenamefont {Jungnitsch}\ \emph {et~al.}(2011)\citenamefont
  {Jungnitsch}, \citenamefont {Moroder},\ and\ \citenamefont
  {G\"uhne}}]{guhne11}%
  \BibitemOpen
  \bibfield  {author} {\bibinfo {author} {\bibfnamefont {B.}~\bibnamefont
  {Jungnitsch}}, \bibinfo {author} {\bibfnamefont {T.}~\bibnamefont {Moroder}},
  \ and\ \bibinfo {author} {\bibfnamefont {O.}~\bibnamefont {G\"uhne}},\ }\href
  {\doibase 10.1103/PhysRevLett.106.190502} {\bibfield  {journal} {\bibinfo
  {journal} {Phys. Rev. Lett.}\ }\textbf {\bibinfo {volume} {106}},\ \bibinfo
  {pages} {190502} (\bibinfo {year} {2011})}\BibitemShut {NoStop}%
\bibitem [{\citenamefont {Lancien}\ \emph {et~al.}(2015)\citenamefont
  {Lancien}, \citenamefont {Gühne}, \citenamefont {Sengupta},\ and\
  \citenamefont {Huber}}]{guhne15}%
  \BibitemOpen
  \bibfield  {author} {\bibinfo {author} {\bibfnamefont {C.}~\bibnamefont
  {Lancien}}, \bibinfo {author} {\bibfnamefont {O.}~\bibnamefont {Gühne}},
  \bibinfo {author} {\bibfnamefont {R.}~\bibnamefont {Sengupta}}, \ and\
  \bibinfo {author} {\bibfnamefont {M.}~\bibnamefont {Huber}},\ }\href
  {\doibase 10.1088/1751-8113/48/50/505302} {\bibfield  {journal} {\bibinfo
  {journal} {Journal of Physics A: Mathematical and Theoretical}\ }\textbf
  {\bibinfo {volume} {48}},\ \bibinfo {pages} {505302} (\bibinfo {year}
  {2015})}\BibitemShut {NoStop}%
\bibitem [{\citenamefont {Huber}\ and\ \citenamefont
  {Sengupta}(2014)}]{sengupta14}%
  \BibitemOpen
  \bibfield  {author} {\bibinfo {author} {\bibfnamefont {M.}~\bibnamefont
  {Huber}}\ and\ \bibinfo {author} {\bibfnamefont {R.}~\bibnamefont
  {Sengupta}},\ }\href {\doibase 10.1103/PhysRevLett.113.100501} {\bibfield
  {journal} {\bibinfo  {journal} {Phys. Rev. Lett.}\ }\textbf {\bibinfo
  {volume} {113}},\ \bibinfo {pages} {100501} (\bibinfo {year}
  {2014})}\BibitemShut {NoStop}%
\bibitem [{\citenamefont {Clivaz}\ \emph {et~al.}(2017)\citenamefont {Clivaz},
  \citenamefont {Huber}, \citenamefont {Lami},\ and\ \citenamefont
  {Murta}}]{huber17}%
  \BibitemOpen
  \bibfield  {author} {\bibinfo {author} {\bibfnamefont {F.}~\bibnamefont
  {Clivaz}}, \bibinfo {author} {\bibfnamefont {M.}~\bibnamefont {Huber}},
  \bibinfo {author} {\bibfnamefont {L.}~\bibnamefont {Lami}}, \ and\ \bibinfo
  {author} {\bibfnamefont {G.}~\bibnamefont {Murta}},\ }\href {\doibase
  10.1063/1.4998433} {\bibfield  {journal} {\bibinfo  {journal} {Journal of
  Mathematical Physics}\ }\textbf {\bibinfo {volume} {58}},\ \bibinfo {pages}
  {082201} (\bibinfo {year} {2017})},\ \Eprint
  {http://arxiv.org/abs/1609.08126} {arXiv:1609.08126} \BibitemShut {NoStop}%
\bibitem [{\citenamefont {Alicki}\ and\ \citenamefont {Lendi}(2007)}]{alicki}%
  \BibitemOpen
  \bibfield  {author} {\bibinfo {author} {\bibfnamefont {R.}~\bibnamefont
  {Alicki}}\ and\ \bibinfo {author} {\bibfnamefont {K.}~\bibnamefont {Lendi}},\
  }\href@noop {} {\emph {\bibinfo {title} {Quantum Dynamical Semigroups and
  Applications}}},\ Lecture notes in Physics\ (\bibinfo  {publisher}
  {Springer-Verlag Berlin Heidelberg},\ \bibinfo {year} {2007})\BibitemShut
  {NoStop}%
\bibitem [{\citenamefont {Breuer}\ and\ \citenamefont
  {Petruccione}(2002)}]{breuer}%
  \BibitemOpen
  \bibfield  {author} {\bibinfo {author} {\bibfnamefont {H.~P.}\ \bibnamefont
  {Breuer}}\ and\ \bibinfo {author} {\bibfnamefont {F.}~\bibnamefont
  {Petruccione}},\ }\href@noop {} {\emph {\bibinfo {title} {The theory of open
  quantum systems}}}\ (\bibinfo  {publisher} {Oxford University Press},\
  \bibinfo {address} {Great Clarendon Street},\ \bibinfo {year}
  {2002})\BibitemShut {NoStop}%
\bibitem [{\citenamefont {Rivas}\ \emph {et~al.}(2014)\citenamefont {Rivas},
  \citenamefont {Huelga},\ and\ \citenamefont {Plenio}}]{rivas1}%
  \BibitemOpen
  \bibfield  {author} {\bibinfo {author} {\bibfnamefont {A.}~\bibnamefont
  {Rivas}}, \bibinfo {author} {\bibfnamefont {S.~F.}\ \bibnamefont {Huelga}}, \
  and\ \bibinfo {author} {\bibfnamefont {M.~B.}\ \bibnamefont {Plenio}},\
  }\href {http://stacks.iop.org/0034-4885/77/i=9/a=094001} {\bibfield
  {journal} {\bibinfo  {journal} {Reports on Progress in Physics}\ }\textbf
  {\bibinfo {volume} {77}},\ \bibinfo {pages} {094001} (\bibinfo {year}
  {2014})}\BibitemShut {NoStop}%
\bibitem [{\citenamefont {Breuer}\ \emph {et~al.}(2016)\citenamefont {Breuer},
  \citenamefont {Laine}, \citenamefont {Piilo},\ and\ \citenamefont
  {Vacchini}}]{breuerN}%
  \BibitemOpen
  \bibfield  {author} {\bibinfo {author} {\bibfnamefont {H.-P.}\ \bibnamefont
  {Breuer}}, \bibinfo {author} {\bibfnamefont {E.-M.}\ \bibnamefont {Laine}},
  \bibinfo {author} {\bibfnamefont {J.}~\bibnamefont {Piilo}}, \ and\ \bibinfo
  {author} {\bibfnamefont {B.}~\bibnamefont {Vacchini}},\ }\href {\doibase
  10.1103/RevModPhys.88.021002} {\bibfield  {journal} {\bibinfo  {journal}
  {Rev. Mod. Phys.}\ }\textbf {\bibinfo {volume} {88}},\ \bibinfo {pages}
  {021002} (\bibinfo {year} {2016})}\BibitemShut {NoStop}%
\bibitem [{\citenamefont {de~Vega}\ and\ \citenamefont
  {Alonso}(2017)}]{alonso}%
  \BibitemOpen
  \bibfield  {author} {\bibinfo {author} {\bibfnamefont {I.}~\bibnamefont
  {de~Vega}}\ and\ \bibinfo {author} {\bibfnamefont {D.}~\bibnamefont
  {Alonso}},\ }\href {\doibase 10.1103/RevModPhys.89.015001} {\bibfield
  {journal} {\bibinfo  {journal} {Rev. Mod. Phys.}\ }\textbf {\bibinfo {volume}
  {89}},\ \bibinfo {pages} {015001} (\bibinfo {year} {2017})}\BibitemShut
  {NoStop}%
\bibitem [{\citenamefont {Bellomo}\ \emph {et~al.}(2007)\citenamefont
  {Bellomo}, \citenamefont {Lo~Franco},\ and\ \citenamefont
  {Compagno}}]{bellomo}%
  \BibitemOpen
  \bibfield  {author} {\bibinfo {author} {\bibfnamefont {B.}~\bibnamefont
  {Bellomo}}, \bibinfo {author} {\bibfnamefont {R.}~\bibnamefont {Lo~Franco}},
  \ and\ \bibinfo {author} {\bibfnamefont {G.}~\bibnamefont {Compagno}},\
  }\href {\doibase 10.1103/PhysRevLett.99.160502} {\bibfield  {journal}
  {\bibinfo  {journal} {Phys. Rev. Lett.}\ }\textbf {\bibinfo {volume} {99}},\
  \bibinfo {pages} {160502} (\bibinfo {year} {2007})}\BibitemShut {NoStop}%
\bibitem [{\citenamefont {Dijkstra}\ and\ \citenamefont
  {Tanimura}(2010)}]{arend}%
  \BibitemOpen
  \bibfield  {author} {\bibinfo {author} {\bibfnamefont {A.~G.}\ \bibnamefont
  {Dijkstra}}\ and\ \bibinfo {author} {\bibfnamefont {Y.}~\bibnamefont
  {Tanimura}},\ }\href {\doibase 10.1103/PhysRevLett.104.250401} {\bibfield
  {journal} {\bibinfo  {journal} {Phys. Rev. Lett.}\ }\textbf {\bibinfo
  {volume} {104}},\ \bibinfo {pages} {250401} (\bibinfo {year}
  {2010})}\BibitemShut {NoStop}%
\bibitem [{\citenamefont {Bhattacharya}\ \emph {et~al.}(2017)\citenamefont
  {Bhattacharya}, \citenamefont {Misra}, \citenamefont {Mukhopadhyay},\ and\
  \citenamefont {Pati}}]{samya1}%
  \BibitemOpen
  \bibfield  {author} {\bibinfo {author} {\bibfnamefont {S.}~\bibnamefont
  {Bhattacharya}}, \bibinfo {author} {\bibfnamefont {A.}~\bibnamefont {Misra}},
  \bibinfo {author} {\bibfnamefont {C.}~\bibnamefont {Mukhopadhyay}}, \ and\
  \bibinfo {author} {\bibfnamefont {A.~K.}\ \bibnamefont {Pati}},\ }\href
  {\doibase 10.1103/PhysRevA.95.012122} {\bibfield  {journal} {\bibinfo
  {journal} {Phys. Rev. A}\ }\textbf {\bibinfo {volume} {95}},\ \bibinfo
  {pages} {012122} (\bibinfo {year} {2017})}\BibitemShut {NoStop}%
\bibitem [{\citenamefont {Mukhopadhyay}\ \emph {et~al.}(2017)\citenamefont
  {Mukhopadhyay}, \citenamefont {Bhattacharya}, \citenamefont {Misra},\ and\
  \citenamefont {Pati}}]{samya2}%
  \BibitemOpen
  \bibfield  {author} {\bibinfo {author} {\bibfnamefont {C.}~\bibnamefont
  {Mukhopadhyay}}, \bibinfo {author} {\bibfnamefont {S.}~\bibnamefont
  {Bhattacharya}}, \bibinfo {author} {\bibfnamefont {A.}~\bibnamefont {Misra}},
  \ and\ \bibinfo {author} {\bibfnamefont {A.~K.}\ \bibnamefont {Pati}},\
  }\href {\doibase 10.1103/PhysRevA.96.052125} {\bibfield  {journal} {\bibinfo
  {journal} {Phys. Rev. A}\ }\textbf {\bibinfo {volume} {96}},\ \bibinfo
  {pages} {052125} (\bibinfo {year} {2017})}\BibitemShut {NoStop}%
\bibitem [{\citenamefont {Awasthi}\ \emph {et~al.}(2018)\citenamefont
  {Awasthi}, \citenamefont {Bhattacharya}, \citenamefont {Sen(De)},\ and\
  \citenamefont {Sen}}]{samya3}%
  \BibitemOpen
  \bibfield  {author} {\bibinfo {author} {\bibfnamefont {N.}~\bibnamefont
  {Awasthi}}, \bibinfo {author} {\bibfnamefont {S.}~\bibnamefont
  {Bhattacharya}}, \bibinfo {author} {\bibfnamefont {A.}~\bibnamefont
  {Sen(De)}}, \ and\ \bibinfo {author} {\bibfnamefont {U.}~\bibnamefont
  {Sen}},\ }\href {\doibase 10.1103/PhysRevA.97.032103} {\bibfield  {journal}
  {\bibinfo  {journal} {Phys. Rev. A}\ }\textbf {\bibinfo {volume} {97}},\
  \bibinfo {pages} {032103} (\bibinfo {year} {2018})}\BibitemShut {NoStop}%
\bibitem [{\citenamefont {Wolf}\ \emph {et~al.}(2008)\citenamefont {Wolf},
  \citenamefont {Eisert}, \citenamefont {Cubitt},\ and\ \citenamefont
  {Cirac}}]{wolf1}%
  \BibitemOpen
  \bibfield  {author} {\bibinfo {author} {\bibfnamefont {M.~M.}\ \bibnamefont
  {Wolf}}, \bibinfo {author} {\bibfnamefont {J.}~\bibnamefont {Eisert}},
  \bibinfo {author} {\bibfnamefont {T.~S.}\ \bibnamefont {Cubitt}}, \ and\
  \bibinfo {author} {\bibfnamefont {J.~I.}\ \bibnamefont {Cirac}},\ }\href
  {\doibase 10.1103/PhysRevLett.101.150402} {\bibfield  {journal} {\bibinfo
  {journal} {Phys. Rev. Lett.}\ }\textbf {\bibinfo {volume} {101}},\ \bibinfo
  {pages} {150402} (\bibinfo {year} {2008})}\BibitemShut {NoStop}%
\bibitem [{\citenamefont {Lu}\ \emph {et~al.}(2010)\citenamefont {Lu},
  \citenamefont {Wang},\ and\ \citenamefont {Sun}}]{nm1}%
  \BibitemOpen
  \bibfield  {author} {\bibinfo {author} {\bibfnamefont {X.-M.}\ \bibnamefont
  {Lu}}, \bibinfo {author} {\bibfnamefont {X.}~\bibnamefont {Wang}}, \ and\
  \bibinfo {author} {\bibfnamefont {C.~P.}\ \bibnamefont {Sun}},\ }\href
  {\doibase 10.1103/PhysRevA.82.042103} {\bibfield  {journal} {\bibinfo
  {journal} {Phys. Rev. A}\ }\textbf {\bibinfo {volume} {82}},\ \bibinfo
  {pages} {042103} (\bibinfo {year} {2010})}\BibitemShut {NoStop}%
\bibitem [{\citenamefont {Rajagopal}\ \emph {et~al.}(2010)\citenamefont
  {Rajagopal}, \citenamefont {Usha~Devi},\ and\ \citenamefont {Rendell}}]{nm2}%
  \BibitemOpen
  \bibfield  {author} {\bibinfo {author} {\bibfnamefont {A.~K.}\ \bibnamefont
  {Rajagopal}}, \bibinfo {author} {\bibfnamefont {A.~R.}\ \bibnamefont
  {Usha~Devi}}, \ and\ \bibinfo {author} {\bibfnamefont {R.~W.}\ \bibnamefont
  {Rendell}},\ }\href {\doibase 10.1103/PhysRevA.82.042107} {\bibfield
  {journal} {\bibinfo  {journal} {Phys. Rev. A}\ }\textbf {\bibinfo {volume}
  {82}},\ \bibinfo {pages} {042107} (\bibinfo {year} {2010})}\BibitemShut
  {NoStop}%
\bibitem [{\citenamefont {Luo}\ \emph {et~al.}(2012)\citenamefont {Luo},
  \citenamefont {Fu},\ and\ \citenamefont {Song}}]{nm3}%
  \BibitemOpen
  \bibfield  {author} {\bibinfo {author} {\bibfnamefont {S.}~\bibnamefont
  {Luo}}, \bibinfo {author} {\bibfnamefont {S.}~\bibnamefont {Fu}}, \ and\
  \bibinfo {author} {\bibfnamefont {H.}~\bibnamefont {Song}},\ }\href {\doibase
  10.1103/PhysRevA.86.044101} {\bibfield  {journal} {\bibinfo  {journal} {Phys.
  Rev. A}\ }\textbf {\bibinfo {volume} {86}},\ \bibinfo {pages} {044101}
  (\bibinfo {year} {2012})}\BibitemShut {NoStop}%
\bibitem [{\citenamefont {Jiang}\ and\ \citenamefont {Luo}(2013)}]{nm4}%
  \BibitemOpen
  \bibfield  {author} {\bibinfo {author} {\bibfnamefont {M.}~\bibnamefont
  {Jiang}}\ and\ \bibinfo {author} {\bibfnamefont {S.}~\bibnamefont {Luo}},\
  }\href {\doibase 10.1103/PhysRevA.88.034101} {\bibfield  {journal} {\bibinfo
  {journal} {Phys. Rev. A}\ }\textbf {\bibinfo {volume} {88}},\ \bibinfo
  {pages} {034101} (\bibinfo {year} {2013})}\BibitemShut {NoStop}%
\bibitem [{\citenamefont {Lorenzo}\ \emph {et~al.}(2013)\citenamefont
  {Lorenzo}, \citenamefont {Plastina},\ and\ \citenamefont
  {Paternostro}}]{nm5}%
  \BibitemOpen
  \bibfield  {author} {\bibinfo {author} {\bibfnamefont {S.}~\bibnamefont
  {Lorenzo}}, \bibinfo {author} {\bibfnamefont {F.}~\bibnamefont {Plastina}}, \
  and\ \bibinfo {author} {\bibfnamefont {M.}~\bibnamefont {Paternostro}},\
  }\href {\doibase 10.1103/PhysRevA.88.020102} {\bibfield  {journal} {\bibinfo
  {journal} {Phys. Rev. A}\ }\textbf {\bibinfo {volume} {88}},\ \bibinfo
  {pages} {020102} (\bibinfo {year} {2013})}\BibitemShut {NoStop}%
\bibitem [{\citenamefont {Dhar}\ \emph {et~al.}(2015)\citenamefont {Dhar},
  \citenamefont {Bera},\ and\ \citenamefont {Adesso}}]{nm6}%
  \BibitemOpen
  \bibfield  {author} {\bibinfo {author} {\bibfnamefont {H.~S.}\ \bibnamefont
  {Dhar}}, \bibinfo {author} {\bibfnamefont {M.~N.}\ \bibnamefont {Bera}}, \
  and\ \bibinfo {author} {\bibfnamefont {G.}~\bibnamefont {Adesso}},\ }\href
  {\doibase 10.1103/PhysRevA.91.032115} {\bibfield  {journal} {\bibinfo
  {journal} {Phys. Rev. A}\ }\textbf {\bibinfo {volume} {91}},\ \bibinfo
  {pages} {032115} (\bibinfo {year} {2015})}\BibitemShut {NoStop}%
\bibitem [{\citenamefont {{Bhattacharya}}\ \emph {et~al.}(2018)\citenamefont
  {{Bhattacharya}}, \citenamefont {{Bhattacharya}},\ and\ \citenamefont
  {{Majumdar}}}]{samya4}%
  \BibitemOpen
  \bibfield  {author} {\bibinfo {author} {\bibfnamefont {S.}~\bibnamefont
  {{Bhattacharya}}}, \bibinfo {author} {\bibfnamefont {B.}~\bibnamefont
  {{Bhattacharya}}}, \ and\ \bibinfo {author} {\bibfnamefont {A.~S.}\
  \bibnamefont {{Majumdar}}},\ }\href@noop {} {\bibfield  {journal} {\bibinfo
  {journal} {e-print}\ } (\bibinfo {year} {2018})},\ \Eprint
  {http://arxiv.org/abs/1803.06881} {arXiv:1803.06881 [quant-ph]} \BibitemShut
  {NoStop}%
\bibitem [{\citenamefont {{Bhattacharya}}\ and\ \citenamefont
  {{Bhattacharya}}(2018)}]{samya5}%
  \BibitemOpen
  \bibfield  {author} {\bibinfo {author} {\bibfnamefont {B.}~\bibnamefont
  {{Bhattacharya}}}\ and\ \bibinfo {author} {\bibfnamefont {S.}~\bibnamefont
  {{Bhattacharya}}},\ }\href@noop {} {\bibfield  {journal} {\bibinfo  {journal}
  {e-print}\ } (\bibinfo {year} {2018})},\ \Eprint
  {http://arxiv.org/abs/1805.11418} {arXiv:1805.11418 [quant-ph]} \BibitemShut
  {NoStop}%
\bibitem [{\citenamefont {{Maity}}\ \emph {et~al.}(2019)\citenamefont
  {{Maity}}, \citenamefont {{Bhattacharya}},\ and\ \citenamefont
  {{Maujmdar}}}]{samya6}%
  \BibitemOpen
  \bibfield  {author} {\bibinfo {author} {\bibfnamefont {A.~G.}\ \bibnamefont
  {{Maity}}}, \bibinfo {author} {\bibfnamefont {S.}~\bibnamefont
  {{Bhattacharya}}}, \ and\ \bibinfo {author} {\bibfnamefont {A.~S.}\
  \bibnamefont {{Maujmdar}}},\ }\href@noop {} {\bibfield  {journal} {\bibinfo
  {journal} {e-print}\ } (\bibinfo {year} {2019})},\ \Eprint
  {http://arxiv.org/abs/1901.02372} {arXiv:1901.02372 [quant-ph]} \BibitemShut
  {NoStop}%
\bibitem [{\citenamefont {Chanda}\ and\ \citenamefont
  {Bhattacharya}(2016)}]{samya7}%
  \BibitemOpen
  \bibfield  {author} {\bibinfo {author} {\bibfnamefont {T.}~\bibnamefont
  {Chanda}}\ and\ \bibinfo {author} {\bibfnamefont {S.}~\bibnamefont
  {Bhattacharya}},\ }\href {\doibase https://doi.org/10.1016/j.aop.2016.01.004}
  {\bibfield  {journal} {\bibinfo  {journal} {Annals of Physics}\ }\textbf
  {\bibinfo {volume} {366}},\ \bibinfo {pages} {1 } (\bibinfo {year}
  {2016})}\BibitemShut {NoStop}%
\bibitem [{\citenamefont {Hall}\ \emph {et~al.}(2014)\citenamefont {Hall},
  \citenamefont {Cresser}, \citenamefont {Li},\ and\ \citenamefont
  {Andersson}}]{hall14}%
  \BibitemOpen
  \bibfield  {author} {\bibinfo {author} {\bibfnamefont {M.~J.~W.}\
  \bibnamefont {Hall}}, \bibinfo {author} {\bibfnamefont {J.~D.}\ \bibnamefont
  {Cresser}}, \bibinfo {author} {\bibfnamefont {L.}~\bibnamefont {Li}}, \ and\
  \bibinfo {author} {\bibfnamefont {E.}~\bibnamefont {Andersson}},\ }\href
  {\doibase 10.1103/PhysRevA.89.042120} {\bibfield  {journal} {\bibinfo
  {journal} {Phys. Rev. A}\ }\textbf {\bibinfo {volume} {89}},\ \bibinfo
  {pages} {042120} (\bibinfo {year} {2014})}\BibitemShut {NoStop}%
\bibitem [{\citenamefont {{Lindblad}}(1976)}]{lindblad}%
  \BibitemOpen
  \bibfield  {author} {\bibinfo {author} {\bibfnamefont {G.}~\bibnamefont
  {{Lindblad}}},\ }\href {\doibase 10.1007/BF01608499} {\bibfield  {journal}
  {\bibinfo  {journal} {Communications in Mathematical Physics}\ }\textbf
  {\bibinfo {volume} {48}},\ \bibinfo {pages} {119} (\bibinfo {year}
  {1976})}\BibitemShut {NoStop}%
\bibitem [{\citenamefont {{Gorini}}\ \emph {et~al.}(1976)\citenamefont
  {{Gorini}}, \citenamefont {{Kossakowski}},\ and\ \citenamefont
  {{Sudarshan}}}]{gorini}%
  \BibitemOpen
  \bibfield  {author} {\bibinfo {author} {\bibfnamefont {V.}~\bibnamefont
  {{Gorini}}}, \bibinfo {author} {\bibfnamefont {A.}~\bibnamefont
  {{Kossakowski}}}, \ and\ \bibinfo {author} {\bibfnamefont {E.~C.~G.}\
  \bibnamefont {{Sudarshan}}},\ }\href {\doibase 10.1063/1.522979} {\bibfield
  {journal} {\bibinfo  {journal} {Journal of Mathematical Physics}\ }\textbf
  {\bibinfo {volume} {17}},\ \bibinfo {pages} {821} (\bibinfo {year}
  {1976})}\BibitemShut {NoStop}%
\bibitem [{\citenamefont {Choi}(1975)}]{choi75}%
  \BibitemOpen
  \bibfield  {author} {\bibinfo {author} {\bibfnamefont {M.-D.}\ \bibnamefont
  {Choi}},\ }\href {\doibase https://doi.org/10.1016/0024-3795(75)90075-0}
  {\bibfield  {journal} {\bibinfo  {journal} {Linear Algebra and its
  Applications}\ }\textbf {\bibinfo {volume} {10}},\ \bibinfo {pages} {285}
  (\bibinfo {year} {1975})}\BibitemShut {NoStop}%
\bibitem [{\citenamefont {Jamiołkowski}(1972)}]{jamil72}%
  \BibitemOpen
  \bibfield  {author} {\bibinfo {author} {\bibfnamefont {A.}~\bibnamefont
  {Jamiołkowski}},\ }\href {\doibase
  https://doi.org/10.1016/0034-4877(72)90011-0} {\bibfield  {journal} {\bibinfo
   {journal} {Reports on Mathematical Physics}\ }\textbf {\bibinfo {volume}
  {3}},\ \bibinfo {pages} {275} (\bibinfo {year} {1972})}\BibitemShut {NoStop}%
\end{thebibliography}%

\end{document}